\documentclass[aps,prl,twocolumn,showpacs,superscriptaddress,floatfix,nofootinbib]{revtex4}

\usepackage{latexsym}
\usepackage{graphicx}
\usepackage{epsfig}
\usepackage[english]{babel}

\begin{document}

\title{Weakly-entangled states are dense and robust}

\author{Rom\'an Or\'us}
\author{Rolf Tarrach}
\affiliation{Dept. d'Estructura i Constituents de la Mat\` eria,
Univ. Barcelona, 08028, Barcelona, Spain.}
\date{\today}

\begin{abstract}
Motivated by the mathematical definition of entanglement we undertake 
a rigorous analysis of the separability and non-distillability properties in the
neighborhood of those three-qubit mixed states which are entangled and completely
bi-separable. Our results are not only restricted to this class of quantum
states, since they rest upon very general properties of mixed states and
Unextendible Product Bases for any possible number of parties. Robustness
against noise of the relevant properties of these states implies the significance of
their possible experimental realization, therefore being of physical -and not
exclusively mathematical- interest.    

\end{abstract}

\maketitle

Quantum mechanics has focused much of the attention of the
physics community during the last century, both for its
extraordinary predictive power and its apparent paradoxes. It has been a
turning point in our conception of the fundamental laws of nature, and it
still leads today to important discussion about its foundations. A major step forward
was realizing that the laws of quantum mechanics can be of practical use. Exploiting 
quantum correlations -or \emph{entanglement}- \cite{einstein, schro} as a fundamental resource (such
as energy) has been proven to be an outstanding success with applications such as 
quantum teleportation, quantum cryptography or quantum computation
\cite{nielsen}. Characterization of entanglement is therefore one of the most relevant
problems in quantum information science. 

It was initially Werner \cite{werner} who
introduced the current mathematical definition of ``mixed entangled state'':
a mixed state $\rho$ of $n$ parties is entangled if and only if it can not be decomposed
as $\rho = \sum_{i=1}^r {\rm p}_i \rho_i^1 \otimes \cdots \otimes \rho_i^n$,
$\{ {\rm p}_i \}_{i=1}^r$ being a certain probability distribution. According to
this definition, any quantum state $\rho$ that is not entangled can always be
created by means of local operations on each separate party together with
classical communication between them (LOCC). A different (but related) concept, the 
\emph{distillability} of quantum states, was introduced by Bennett \emph{et
al.} \cite{bennett1}: given $M$ copies of a bipartite quantum state $\rho$,
we say that it is distillable if and only if we can obtain $N$ copies of a
maximally entangled pure state of the two parties by means of LOCC (the
generalization to the $n$-partite case is straightforward, by separating the $n$ parties into
two different sets, and applying again the same definition). This naturally
led to the concept of \emph{distillable entanglement} of state $\rho$, which corresponds to
the ratio $N/M$ in the limit of infinite number of copies. 

The existence of
entangled quantum states that are non-distillable, the so-called
\emph{bound entangled} states, was soon proved by Horodecki \emph{et al.}
\cite{horodecki}. The discovered states had positive partial
transposition (PPT) with respect to one of the two subsystems considered for
the distillation protocol. Despite that recent results have found 
particular practical applications for some of these states
\cite{horo2, cubitt, acin}, 
it still seems that most of them (e.g. those mixed states
of three qubits which, despite being entangled, are separable with respect
to all the possible bi-partitions of the system \cite{bennett, upb2})
are actually so weakly entangled that their quantum correlations are not known
to be of any 
practical use for quantum information tasks. This situation
leads us to think that, perhaps, our definition of entanglement is excessively 
mathematical, and that a more physical definition of what a
quantum-correlated state is, might be of interest. The importance of this
problem hinges on whether it is actually possible to realize bound entangled
states, which posses this ``useless'' character, in the laboratory.
For this to be achievable we 
must demand as a necessary condition that the action of
small and unavoidable decoherence effects leave the essential properties of
the quantum state unchanged. If, for instance, an a priori 
interesting property were only to hold for pure states, it would actually not
be of physical interest, as pure states are a mathematical idealization of
physically realizable states. But this experimental condition must be considered
for mixed states in general. 

In this paper we prove that mixed quantum states that are entangled and
are bi-separable with respect to any set of bi-partitions of the system form a set of 
non-zero measure in Hilbert space. In particular this implies that those
three-qubit density matrices that are bound entangled and completely
bi-separable \cite{bennett, upb2} have real physical significance. 
Our results have been obtained by comparing the properties of
the neighborhood in Hilbert space around these states with those of the
original state. Motivated by this class of three-qubit states we
also undertake a more general analysis of the distillability properties of the 
neighborhood of those bound entangled states associated to an Unextendible Product Basis.   

Our study begins by considering the following theorem:

\smallskip

{\bf Theorem 1:} $n$-qubit density matrices $\rho$ which are separable with respect to
some particular bipartition of the system form a set of non-zero measure in Hilbert space
(it is a \emph{dense} set).

\smallskip

{\bf Proof:} we consider a density matrix $\rho$ of $n$ qubits, which is
separable with respect to some particular bi-partition. Let us introduce
independent perturbations in the parameter space of $n$-qubit density matrices
in such a way that the perturbed operator is still separable with respect to that
bi-partition. For this purpose, we consider the following set of $2^n \times
2^n$ independent and completely separable projectors: 
\begin{equation}
E(j_{1}\ldots j_{n}) \equiv |j_{1} \ldots j_{n} \rangle \langle j_{1} \ldots
j_{n}| \ \ \ j_{\alpha} = \{0,1,\phi_1,\phi_2 \} \ , 
\label{projectors}
\end{equation}
for $\alpha = 1, \ldots n$, $|\phi_1 \rangle = \frac{1}{\sqrt{2}} |0 +
1\rangle$ and $|\phi_2 \rangle = \frac{1}{\sqrt{2}} |0 + i 1\rangle$. 
We can now write any perturbation by using this set of
projectors as $\rho(\epsilon_{\mu}) = \frac{1}{C}\left(\rho + \sum_{\mu}
\epsilon_{\mu} E(\mu) \right)$, $\epsilon_{\mu}$ being a set of real 
parameters, $\mu \equiv (j_1 \ldots j_n)$ and $C$ the normalization constant. 
For the perturbed density matrix to be physical
the set of parameters $\epsilon_{\mu}$ must be such that the condition of
positive semi-definiteness holds for the perturbed
operator. Because the projectors $E(\mu)$, which form a basis of the $2^n
\times 2^n$ density matrices, correspond to separable pure states, 
if we restrict ourselves to the region $\epsilon_{\mu} \ge 0 \ \forall \mu$
(which physically corresponds to a local noise), these states share at
least the same separability properties as those of the unperturbed state
$\rho$. It is then proven that these states form a set of non-zero
measure in Hilbert space, since the number of independent perturbation parameters is
maximal, and the separability property is robust against local noise. $\Box$

\smallskip 

Our next claim is that given an entangled state $\rho$, its entanglement is
preserved in an infinitesimal neighborhood. This is proven in our second theorem:

\smallskip

{\bf Theorem 2:} if the $n$-qubit state $\rho$ is entangled and the real
parameters $\epsilon_{\mu}$ are
infinitesimal, then the state $\rho(\epsilon_{\mu}) = \frac{1}{C}\left(\rho + \sum_{\mu}
\epsilon_{\mu} E(\mu) \right)$ is entangled (in particular, entanglement is a 
\emph{robust} property). 

\smallskip

{\bf Proof:} the proof is based on witness operators. Let us briefly
recall their definition: given two convex subsets $S_1$ and $S_2$ such that $S_1$ is
included in $S_2$, a hermitian operator $W$ is a witness operator if and only
if (i) $\forall \sigma \in S_1, \ {\rm tr}(W\sigma) \ge 0$, (ii) there is at
least one $\rho \in S_2$ such that ${\rm tr}(W\rho) < 0$ and (iii) ${\rm
  tr}(W) = 1$. In the quantum case, $S_1$ is the set of separable states,
$S_2$ is the set of all quantum states and $W$ is an observable that has
expectation value $\ge 0$ for all the separable states and $<0$ for some
entangled state. It is a well-known fact \cite{horo3, terhal, dagmar} that a quantum
state $\rho$ is entangled if and only if there exists a witness operator $W$
that ``detects'' $\rho$, i.e. ${\rm tr}(W\rho) < 0$ and ${\rm tr}(W\sigma) \ge 0
\ \forall \sigma$ separable (this is a consequence of the Hahn-Banach's
theorem for convex sets). Given the perturbed state $\rho(\epsilon_{\mu})$, it
will be entangled if and only if there exists a witness operator $W$ such that
${\rm tr}(W\rho(\epsilon_{\mu})) < 0$. Assuming that $\rho$ is entangled, we
choose to work with the particular witness operator $W_{\rho}$ that
``detects'' it (i.e. ${\rm tr}(W_{\rho} \rho) < 0$). For
$\rho(\epsilon_{\mu})$ to be detected by $W_{\rho}$ we impose that 
\begin{equation}
{\rm tr}(W_{\rho} \rho(\epsilon_{\mu})) = \frac{1}{C} \left({\rm tr}(W_{\rho}
\rho) + \sum_{\mu} \epsilon_{\mu} {\rm tr}(W_{\rho} E(\mu))\right) < 0 \ .
\label{condition}
\end{equation}
Because ${\rm tr}(W_{\rho} \rho) < 0$ and ${\rm tr}(W_{\rho}E(\mu))
\ge 0$ (since $E(\mu)$ are projectors corresponding to completely separable
pure states),  the above condition reads $\sum_{\mu} \epsilon_{\mu}
{\rm tr}(W_{\rho}E(\mu)) < | {\rm tr}(W_{\rho} \rho)|$, which can always be
achieved for values of the parameters $\epsilon_{\mu}$ close enough to zero. 
The perturbed state is then detected by some witness operator, therefore 
it is entangled. $\Box$

\smallskip

From the preceding two theorems we infer that there exists a partial neighborhood
of non-zero measure of those $n$ qubit states that
are entangled but separable with respect to some bi-partition that shares also
the same properties, so that these states are robust. In particular, this
result holds for those states of three qubits studied in \cite{tarrach}, and
specifically to the (bound) entangled three-qubit states that are entangled yet
completely bi-separable, which were presented in
\cite{bennett} and further generalized in \cite{bravyi}.
Let us recall at this point their definition in terms of an Unextendible Product Basis: an 
Unextendible Product Basis (UPB) \cite{bennett, upb2} for a multipartite quantum system is an incomplete
orthogonal product basis whose complementary subspace contains no product
state. It has the remarkable property that if $\{ |\psi_i\rangle \}_{i=1}^m$ are the
(product) vectors of the UPB, then the maximally mixed state in the subspace
orthogonal to the UPB, $\rho = \frac{1}{D-m}(I - \sum_{i=1}^m |\psi_i\rangle
\langle \psi_i|)$ (being $D$ the dimensionality of the Hilbert space), is
bound entangled (UPB-states). For three-qubit systems, it was proven in
\cite{bravyi} that the most general UPB is given by the set of
vectors
\begin{eqnarray}
|\psi_1\rangle &=& |0\rangle|0\rangle|0\rangle \nonumber \\
|\psi_2\rangle &=& |1\rangle|B\rangle|C\rangle \nonumber \\ 
|\psi_3\rangle &=& |A\rangle|1\rangle|\bar{C}\rangle \nonumber \\
|\psi_4\rangle &=& |\bar{A}\rangle|\bar{B}\rangle|1\rangle \ ,
\label{upb}
\end{eqnarray} 
with $\langle A|\bar{A}\rangle = 0$ (similar for $|B\rangle$ and $|C\rangle$), and $|A\rangle$,
$|B\rangle$ and $|C\rangle$ depending on only one real parameter each one
(which can always be achieved by a local change of basis). In \cite{bravyi}
it was also noted that any UPB-state of three qubits is bound entangled yet completely
bi-separable. Our theorems
imply that, apart from these UPB-states, there are also mixed density matrices of
this kind which can not be
associated to any UPB of three qubits, since the states $\rho(\epsilon_{\mu})$ obtained by 
perturbing one of the UPB-states are not
in general related to any UPB. Neither can these states always be
written as a convex combination of states associated to some UPB, since a
convex combination of two different UPB-states for three qubits can easily be
seen to have at least rank $6$, while it is possible to achieve states of rank
$5$ simply by mixing a UPB-state with one of the pure product states of the
corresponding UPB. 

We now wish to analyze the distillability properties of the neighborhood of
this kind of three-qubit UPB-states in a more detailed way. Theorem 1 already
implies a certain kind of (obvious) robustness for the non-distillability of these
states, restricted to the $\epsilon_{\mu} \ge 0$ region $\forall \mu$. 
We shall see that, indeed, this robustness is stronger 
due to some peculiarities of UPB-states, as it can also be extended to some cases with
small negative values of the parameters $\epsilon_{\mu}$. Our analysis focuses only on general
properties of UPBs, no matter what the particular system is, therefore its
range of application is not restricted only to the three-qubit case. First we
present a previous lemma which we will use to prove our third theorem:

\smallskip

{\bf Lemma:} if $\rho$ is a UPB-state in ${\mathcal H}_a \otimes
{\mathcal H}_b$, then the kernel (null eigenspace) of $\rho^{T_a}$ 
is spanned by product vectors.

\smallskip

{\bf Proof:} let $\{ |a_i\rangle |b_i\rangle \}_{i = 1}^m$ be a UPB with $m$
product vectors in a Hilbert space ${\mathcal H}_a \otimes {\mathcal H}_b$ of
dimension $D$. The UPB-state is $\rho = \frac{1}{D-m} \left( I - \sum_{i=1}^m 
|a_i\rangle \langle a_i|\otimes |b_i\rangle \langle b_i| \right)$, being its kernel
spanned by the set of vectors of the UPB. Taking the partial transposition with respect to
party $a$ we get $\rho^{T_a} = \frac{1}{D-m} \left( I - \sum_{i=1}^m
(|a_i\rangle \langle a_i|)^{T_a} \otimes |b_i\rangle \langle b_i|
\right)$. Since $|a_i\rangle \langle a_i|$ is a hermitian operator, it holds
that $(|a_i\rangle \langle a_i|)^{T_a} = (|a_i \rangle \langle a_i|)^* = |a'_i
\rangle \langle a'_i|$, being $|a'_i\rangle = (|a_i\rangle)^*$ the
complex-conjugated vector of $|a_i\rangle$. The kernel of $\rho^{T_a}$ is therefore
spanned by the product vectors $\{|a'_i\rangle |b_i\rangle \}_{i=1}^m$. $\Box$

\smallskip

At this point we are in conditions of presenting our third theorem:  

\smallskip

{\bf Theorem 3:} any UPB-state $\rho$ perturbed by a small enough amount of
noise $\rho_1$, such that $\rho_1^{T_a} > 0$ in the kernel of $\rho^{T_a}$, 
remains non-distillable (non-distillability is a conditionally \emph{robust} property).

\smallskip

Before we prove Theorem 3, let us recall that any
physical noise can always be represented in terms of a mixture (with positive weights) 
in the space of density matrices, that is, as a probabilistic combination 
of the original (unperturbed) mixed state and a noise-induced density matrix. 

\smallskip

{\bf Proof:} the proof is
based on degenerate perturbation theory. Let us consider a UPB $\{|a_i\rangle
|b_i\rangle \}_{i=1}^m$ in ${\mathcal H}_a \otimes {\mathcal H}_b$ and call
$\rho$ the corresponding UPB-state. We wish to note here that any UPB can
always be written in this way, by joining the different parties into two
different sets $a$ and $b$. Consider also $\rho_1$ as any other possible
quantum state in the same Hilbert space. We write a small perturbation of
$\rho$ with $\rho_1$ as $\rho(\epsilon) = \frac{1}{1+\epsilon} \left( \rho + \epsilon
\rho_1 \right)$, $\epsilon > 0$ being an infinitesimal noise 
parameter. Taking the partial transposition with respect to one of the parties
we obtain $\rho(\epsilon)^{T_a} = \frac{1}{1+\epsilon} \left(\rho^{T_a} +
\epsilon \rho_1^{T_a} \right)$. The null eigenvectors of $\rho$ can be
chosen to be the states of the UPB $\{|a_i\rangle |b_i\rangle \}_{i=1}^m$, 
while the states from the $(D-m)$-dimensional subspace orthogonal to
this set have eigenvalue $\frac{1}{D-m}$. According to the previous Lemma, the 
kernel of $\rho^{T_a}$ is then spanned by the set of vectors 
$\{ |a'_i\rangle |b_i\rangle \}_{i=1}^m$, while the vectors from its 
orthogonal subspace have eigenvalue $\frac{1}{D-m}$. Using degenerate
perturbation theory, for $\epsilon$ small enough the lowest eigenvalues of
$\rho(\epsilon)^{T_a}$ are given by $\epsilon \lambda_r + O(\epsilon^2)$, 
$\{\lambda_r\}_{r = 1}^m$ being the eigenvalues of the $m \times m$ matrix $A$ defined by
$A_{ij} \equiv \langle a'_i| \langle b_i | \rho_1^{T_a}
|b_j\rangle|a'_j\rangle$. If $\lambda_r < 0$ for some $r$,
this leads to a negative eigenvalue in the spectrum of
$\rho(\epsilon)^{T_a}$, therefore turning $\rho(\epsilon)$ into an
NPT state. The case in which $\lambda_r = 0$ for some $r$ needs a
second-order analysis in perturbation theory, which easily leads to negative eigenvalues in the
spectrum of $\rho(\epsilon)^{T_a}$ as well, and therefore to similar
conclusions. Consequently, in order for the perturbed operator to remain PPT,
we must demand the condition $A > 0$, which means that $\rho_1^{T_a} > 0$ in
the kernel of $\rho^{T_a}$. $\Box$

\smallskip

Note that, in terms of the projectors $E(\mu)$ from Theorem 1, any $\rho_1$ can always be
decomposed as $\rho_1 = \sum_{\mu} c_{\mu} E(\mu)$, being $c_{\mu}$ certain
real parameters. Defining $\epsilon_{\mu}
\equiv \epsilon  c_{\mu}$, we observe that the hypothesis of Theorem 3 does
not restrict $\epsilon_{\mu}$ to be non-negative.
It is also worth to point out the case in which the UPB is composed of
real vectors only. In this situation, the kernel of the partially-transposed
UPB-state $\rho^{T_a}$ coincides with the space spanned by the vectors of
the UPB, and therefore the condition imposed on the noise in Theorem 3 gets 
simplified. This simplification applies to a very large
variety of UPBs, such as all the three-qubit and many two-qutrit examples 
\cite{bennett, upb2, bravyi}.

When particularizing to the three-qubit case, and
bringing together the results from the previous three theorems, we conclude that
mixed three-qubit states that are entangled yet completely bi-separable are
robust against local noise and beyond, in the sense that their entanglement, their complete bi-separability
and their non-distillability are not modified by small local effects and some
non-local noisy effects. Relevant
properties of these states are then of physical significance, and not just a
matter of mathematical interest. We wish to note the dependence of
these states on the existence of UPBs for three qubits, which are associated
to Hilbert spaces of dimension $4$ with no product vectors in them, and we wonder, as a
possible generalization of this concept, whether there
exist subspaces of dimension $5$ of three-qubit Hilbert spaces such that there
are less than $5$ independent product vectors in them (we have not succeeded in
finding them). In such a case, the properties of
these subspaces would probably be of interest, as are the properties of UPBs,
in order to bring further insight and knowledge about entanglement and 
distillability properties for three-partite systems.    

\smallskip

{\bf Acknowledgments:} we are grateful to discussions with A. Ac{\'i}n, 
P. Hyllus, J. I. Latorre, M. Lewenstein and A. Sanpera. We also particularly thank
J. I. Cirac for pointing out an error in a previous version of Theorem 3. 
We acknowledge financial support from projects MCYT FPA2001-3598, GC2001SGR-00065 and
IST-1999-11053.

{}

\end{document}